\documentclass[conference,letterpaper]{IEEEtran}
\usepackage[letterpaper, top=0.7in, bottom=1in, left=0.625in, right=0.625in]{geometry}

\usepackage{cite}

\usepackage{caption, subcaption}

\ifCLASSINFOpdf
  \usepackage[pdftex]{graphicx}
  \graphicspath{{./figures/}}
  \DeclareGraphicsExtensions{.pdf,.jpeg,.png}
\else
\fi
\usepackage{amsmath, amssymb, bm}
\usepackage{xcolor}
\usepackage{soul}
\begin{document}
%


\title{Harmonic Modeling, Data Generation, and Analysis of Power Electronics-Interfaced Residential Loads}

%
%
%
\author{\IEEEauthorblockN{Ankit Singhal,
Dexin Wang, Andrew P. Reiman, Yuan Liu,
Donald J. Hammerstrom, and Soumya Kundu}
\IEEEauthorblockA{Electricity Infrastructure and Buildings Division\\ Pacific Northwest National Laboratory, Richland, WA, USA\\
Email: \{ankit.singhal, dexin.wang, andrew.reiman, yuan.liu, donald.hammerstrom, soumya.kundu\}@pnnl.gov
}}

\maketitle

\begin{abstract}
Integration of electronics-based residential appliances and distributed energy resources in homes is expected to rise with grid decarbonization. These devices may introduce significant harmonics into power networks that need to be closely studied in order to accurately model and forecast load. However, it can be difficult to obtain harmonic-rich voltage and current data -- necessary for identifying accurate load models -- for residential electrical loads. Recognizing this need, first a set of electronics-based end-use loads is identified and modeled in an electromagnetic transients program  tool for a residence. 
Second, an impedance-varying method is proposed to generate harmonic data that captures harmonic propagation to the supply voltage and harmonic interactions among end-use loads connected to the same supply voltage. Third, a harmonic-rich dataset produced via the proposed methodology is demonstrated to successfully identify frequency coupling matrix--based harmonic load models using the least-squares method. Numerical results demonstrate the accuracy of the model. The impact of limited data availability on model identification is also explored. 
\end{abstract}


%
\IEEEpeerreviewmaketitle

\section{Introduction}

Conventional load models, e.g., the constant impedance (\textit{Z}), current (\textit{I}) and power (\textit{P}), or ``ZIP'' models, predominantly assume that the system is operated at the fundamental frequency (50/60 Hz). However, increasing deployment of power electronic devices, such as inverter-based renewable generation and other nonlinear loads (e.g., variable frequency drives; VFDs) contributes to increased harmonic components in system voltages and currents, rendering conventional ZIP models ineffective \cite{mclorn2017enhanced,grady2012understanding}
Enhanced load models are needed to mitigate the effects of harmonics on the power system, which include increased losses, overheated transformers, and melting of capacitor fuses \cite{grady2012understanding,Brunoro2017}. Many of the existing harmonic load models (e.g., scalar multiplier-based correction of ZIP model \cite{mclorn2017enhanced} and current source models \cite{collin_component-based_2010, Salles_2012}) suffer from two main disadvantages. First, in the current source models, harmonic spectra across various residential electrical appliances are required
\cite{collin_component-based_2010, Salles_2012}. With increasing variety in power electronics end-use loads, the frequency spectrum data of various load appliances, typically collected from field measurements and surveys \cite{patidar_harmonics_2009}, may not be publicly available or may not be reliable. Second, these models are voltage-independent
(i.e., neglect the supply voltage harmonics) and cannot capture the real-world phenomenon of \textit{attenuation and diversity effect}, first reported in \cite{mansoor_investigation_1995}. This effect causes the harmonics in supply current to decrease or attenuate with the increasing distortion in supply voltage. Not modeling the harmonics in voltage leads to overestimating the harmonics in the power system \cite{mansoor_investigation_1995, grady_estimating_2002}.


To circumvent this issue, two types of voltage-dependent harmonic load models are developed in the literature: the classical Norton model and the frequency coupling matrix (FCM). Unlike the classical Norton models \cite{zhao2004harmonic}, the FCM and its variants (first proposed in \cite{Fauri_1997}) offer mathematical models to capture the cross-coupling between the voltage and current harmonics of different orders -- a characteristic typical of the power-electronic and nonlinear loads\cite{grady2012understanding}. 
%
Nonetheless, generating a reliable harmonic-rich voltage and current dataset is a crucial and challenging task for identifying FCM. 
While some prior efforts consider deriving closed-form (analytical) expressions for FCM from physics-based device models \cite{lehn2007frequency}, most have resorted to probabilistic computational (\cite{malagon-carvajal_harmonic_2020,lennerhag_stochastic_2020}), and experimental methods \cite{senra2017assessment,Brunoro2017}. 
However, methods that rely on generating extensive experimental datasets in the laboratory by directly varying the harmonic contents of the supply voltage are not applicable in real-world settings. Moreover, distribution network voltage harmonics are affected by other (nonlinear) load currents that need to be reflected in the data as well. 

\textbf{Contributions.} Recognizing this need for data, this work builds a residential home load model by developing electromagnetic transients program (EMTP) models of individual electronic end-use loads with high-frequency switching components connected to the same residential supply. Transient simulation is used to generate a harmonic-rich dataset. This paper proposes a systematic methodology to capture (i) the propagation of harmonic contamination of electronic loads to the home supply voltage and subsequently, (ii) the impact of such distorted supply voltage on the supply currents drawn by the loads (\textit{attenuation effect}). Further, the generated synthetic dataset is used to learn the FCM-based harmonic load models using the \textit{least squares} method. Implications of limited data on identifiability of FCM is also explored. 
%
Sec.\,\ref{sec:methodology} discusses the house modeling, methodology for harmonic-rich data generation, and FCM estimation. Sec.\,\ref{sec:enduse_modeling} presents the end-use load modeling followed by the results in Sec.\,\ref{sec:results}. The conclusion is given in Sec.\,\ref{sec:conclusion}.

\section{Methodology } \label{sec:methodology}

\subsection{System Modeling}
\begin{figure}[]
    \centering
    \includegraphics[width=1.00\columnwidth]{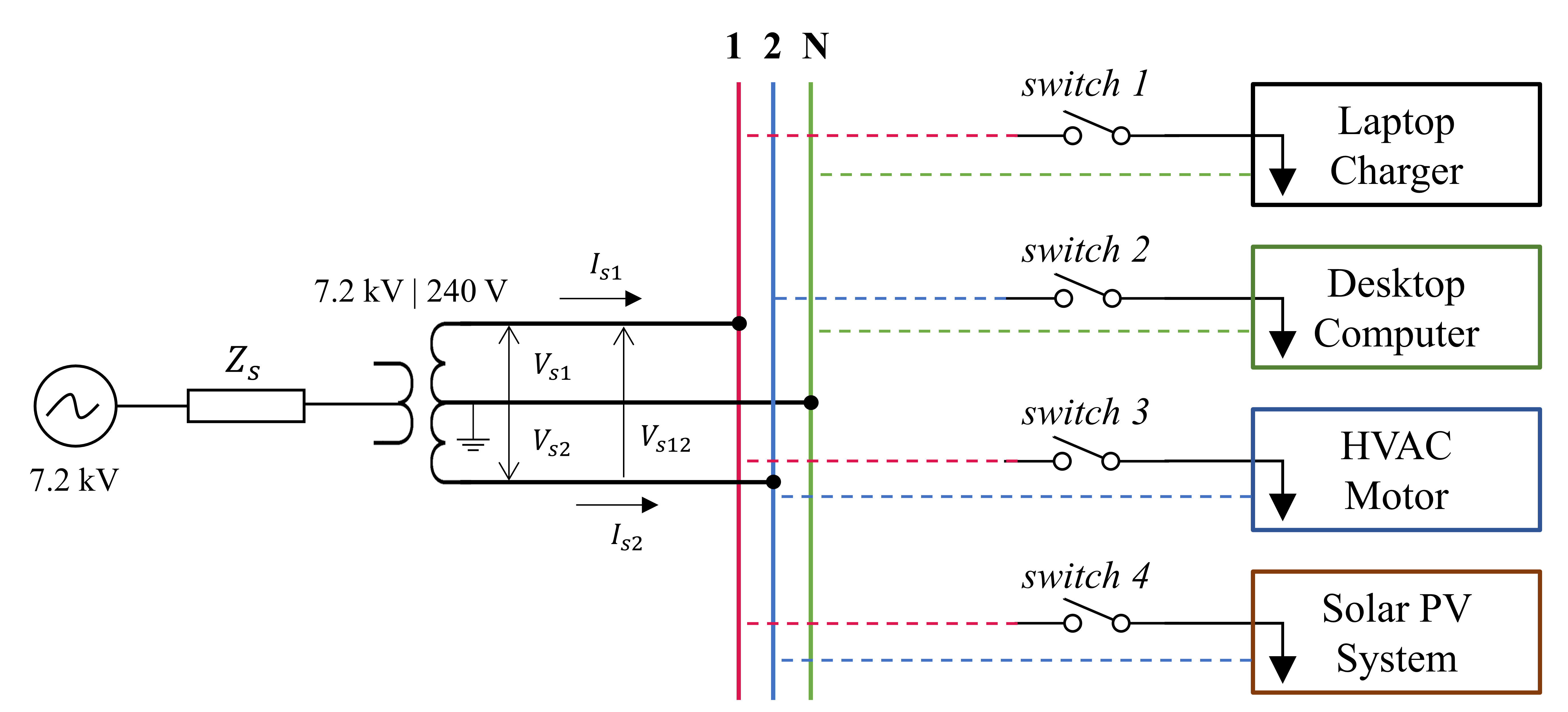}
    \caption{Schematic of a residential house with a split-phase supply and end-use load models}
    \label{fig:overall_ckt}
    \vspace{-0mm}
\end{figure}

\begin{table}
\renewcommand{\arraystretch}{1.2}
\caption{Power electronics based load models representing house appliances }
\label{tab:load_models}
\centering
\begin{tabular}{l l}
\hline
 Load model & House appliances \\
\hline
Rectifier + Buck DC-DC converter & Desktop, home entertainment\\
Rectifier + Flyback DC--DC converter & Laptop charger \\
VFD + Induction motor & HVAC, washer, dryer \\
Boost converter + inverter & PV system, EV charger\\
\hline
\end{tabular}
\vspace{-3mm}
\end{table}


In North America, residential customers are usually supplied through a single-phase split-phase supply consisting of one neutral and two hot phases, $V_{s1}$ and $V_{s2}$, each of 120 V, as shown in \figurename \ref{fig:overall_ckt}. 
The equivalent supply voltage, $V_s^{eq}$, and supply current, $I_s^{eq}$ are defined such that the power measured at the meter remains the same, as follows:
\begin{gather}
    V_s^{eq} = (V_{s1}+V_{s2})/2\,,\,~I_s^{eq} = (V_{s1}\, I_{s1} + V_{s2}\, I_{s2})/V_s^{eq}\,
\end{gather}
The house mains is connected to a 7.2\,kV source via a distribution transformer and line impedance $Z_s$. 
As shown in Fig. \ref{fig:overall_ckt}, the home is modeled as a set of end-use loads, each of which can be individually switched on or off to create various load combinations. %
%
Four power electronics-based end-use loads were modeled, which represent a wide variety of house appliances as shown in Table \ref{tab:load_models}. The desktop and laptop loads are connected to a 120 V supply, whereas the HVAC motor and solar PV systems are connected to the 240 V supply.
Other loads such as resistive heating have less harmonic content, and their contribution to the home load may be modeled more accurately using superposition; therefore, these loads are not modeled explicitly. Individual end-use models are discussed in detail in Sec.\,\ref{sec:enduse_modeling}.

\subsection{Data Generation Methodology}
%
Power electronics based end-uses draw load currents with harmonic content, due to switching controls, introducing harmonics in the supply voltage. Consequently, harmonics in the supply voltage affect the harmonic content of the load current. Thus, the harmonic content of each end-use load affects the harmonic content of other end-use loads (i.e., the home load is not equal to the superposition of independent end-use loads). Furthermore, different combinations of end-use loads lead to different steady-state harmonic load profiles. 
16 possible device combinations are configured using breaker switches (in \figurename \ref{fig:overall_ckt}).

\subsubsection{Modeling Harmonics in Supply Voltage}
The combination of various loads provides us the dataset of harmonic rich current waveforms. We need a varying amount of harmonics in supply voltage and the corresponding supply current to generate sufficient data points for harmonic load modeling. However, the load current of a single residence is not enough to introduce the realistic distortion in the supply voltage. One alternative is to model multiple house loads to obtain an increased load current and consequently distorted supply voltage. However, it makes the EMTP simulation computationally intractable. Therefore, this work adopts the mathematically equivalent alternative of increasing source impedance, $Z_s$, rather than the load current to cause higher distortion in supply voltage with the original magnitude of load current. $Z_s$ is increased by a varying gain factor $\alpha_{imp}$ that introduces harmonics in supply voltage due to the voltage drop effect.

\subsubsection{Overall Process of Data Generation}
\begin{figure}
    \centering
    \includegraphics[width=1\columnwidth]{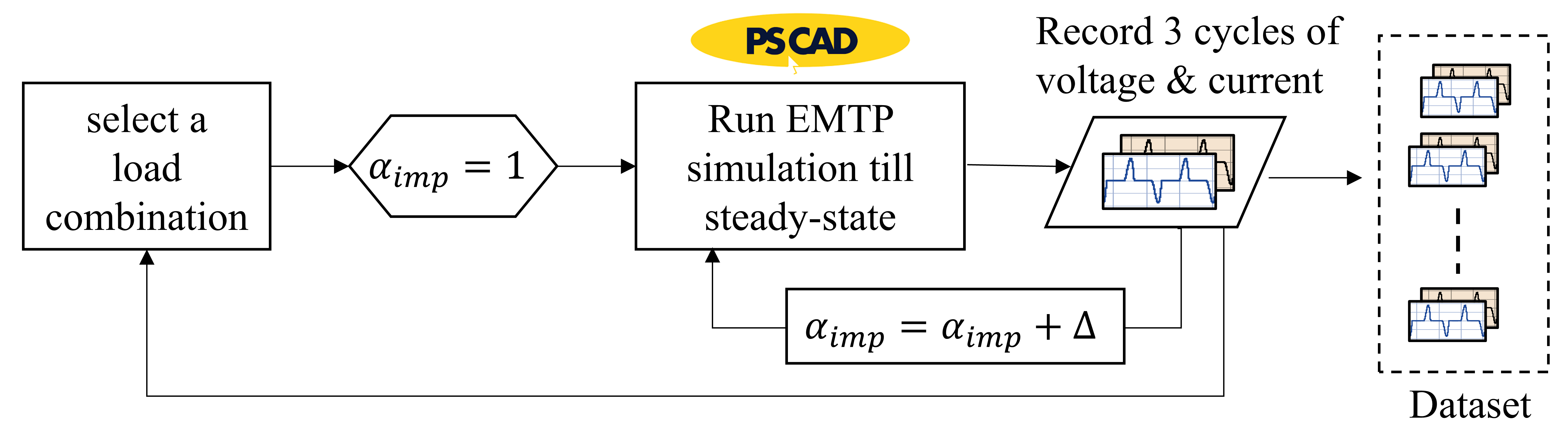}
    \caption{The process of generating harmonic-rich datasets for data-driven harmonic load modeling}
    \label{fig:overall_process}
    \vspace{-4mm}
\end{figure}

\figurename\,\ref{fig:overall_process} shows the overall process of generating a dataset for data-driven load modeling. First, a load combination is selected for which the PSCAD simulation is run until the starting transients are settled for $\alpha_{imp}=1$. The three steady-state cycles of $V_s^{eq}$ and $I_s^{eq}$ waveforms are recorded that constitute one data point. The same process is repeated multiple times for increasing values of $\alpha_{imp}$, and a complete dataset is obtained.

\begin{figure*}[!ht]
\centering
\begin{subfigure}[b]{0.45\linewidth}
\centering
\includegraphics[width=\linewidth]{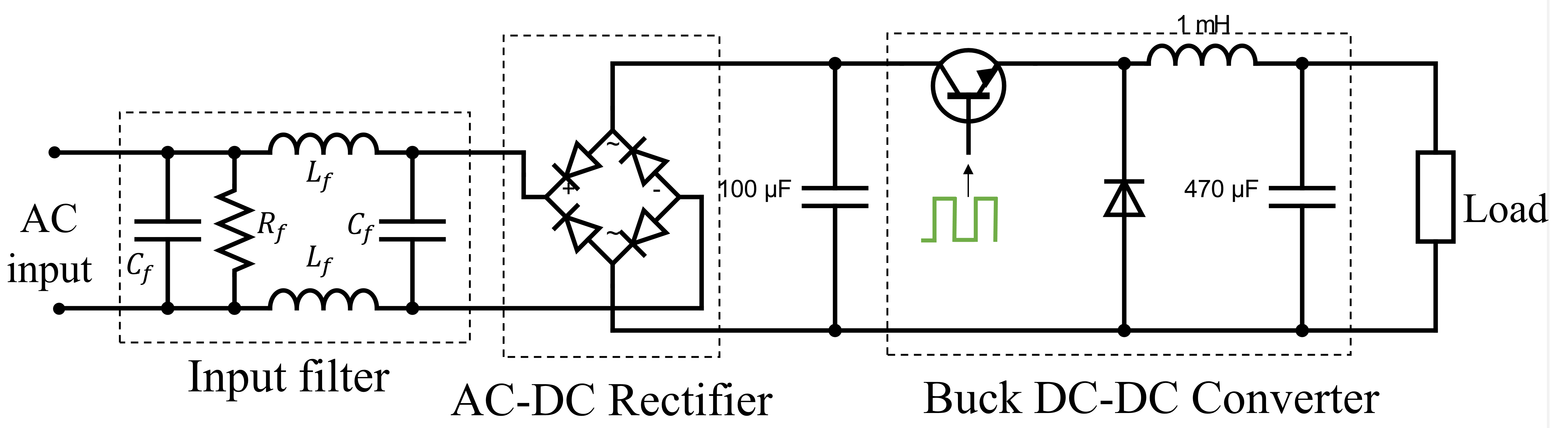}%
\caption{Desktop power supply}
\label{fig:desktop_supply}
\end{subfigure}
\hfil
\begin{subfigure}[b]{0.45\linewidth}
\centering
\includegraphics[width=\linewidth]{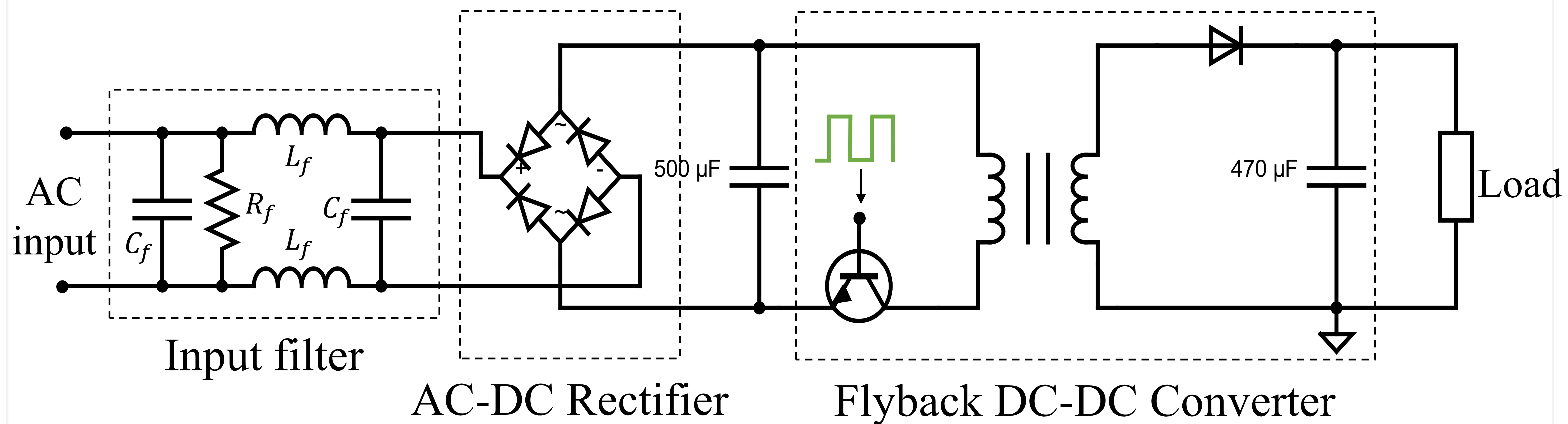}%
\caption{Laptop charger}
\label{fig:laptop_supply}
\end{subfigure}
\hfil
\begin{subfigure}[b]{0.45\linewidth}
\centering
\includegraphics[width=\linewidth]{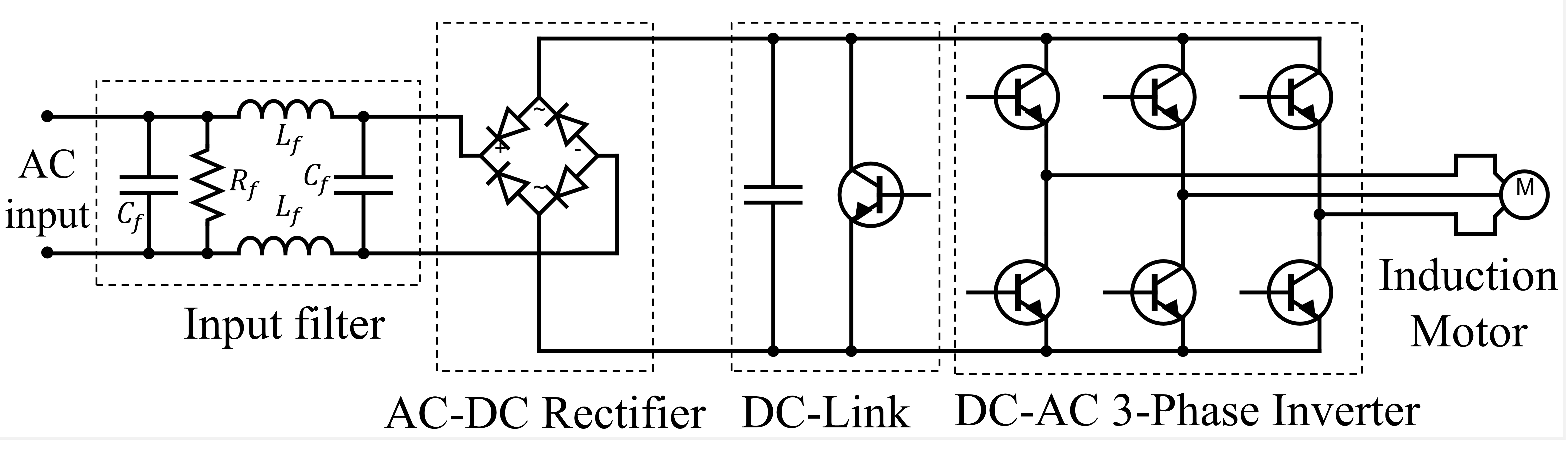}%
\caption{VFD driven induction motor}
\label{fig:vfd}
\end{subfigure}
\hfil
\begin{subfigure}[b]{0.45\linewidth}
\centering
\includegraphics[width=\linewidth]{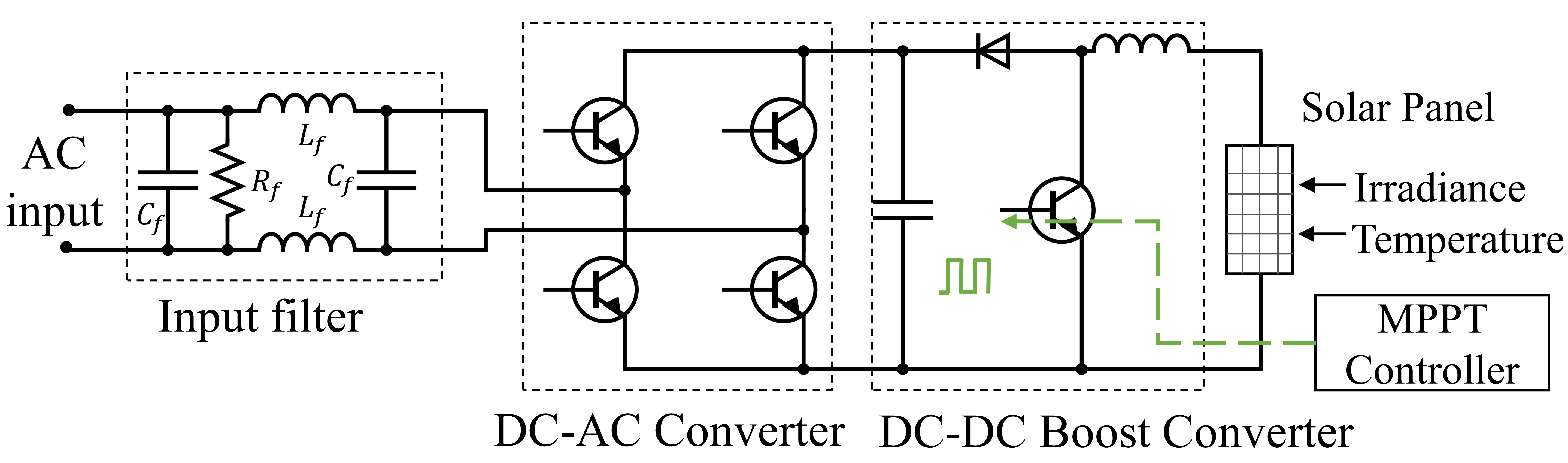}%
\caption{PV System}
\label{fig:pv}
\end{subfigure}
\hfil
\caption{Schematic of various load appliances with input filters modeled in PSCAD}
\vspace{-4mm}
\label{fig:compare_sim_freq}
\end{figure*}
\subsection{FCM: Background and Estimation Methodology}
The FCM, alternatively called the \textit{harmonic admittance matrix} or the \textit{cross-frequency admittance matrix}, was introduced in \cite{Fauri_1997} to represent the coupling between the harmonic components of different orders in the voltage and current signals of the load. In order to construct the FCM, it is assumed that the electrical load is \textit{stationary} (i.e., load response to same voltage signal remains unchanged) and \textit{passive}, and the voltage and current signals are \textit{periodic}. 
Under these assumptions, and allowing for constant (voltage-independent) current injections, the relationship between the current harmonics vector (${\bm i}$) and the voltage harmonics vector (${\bm v}$) is given by:
\begin{align}
    \text{(FCM)}\quad{\bm i}&={\bm i_0} + {\bm Y}{\bm v} = \begin{bmatrix}
    {\bm i_0} & {\bm Y}
    \end{bmatrix} \cdot \begin{bmatrix}
    1 & {\bm v}^\top
    \end{bmatrix}^\top \label{eq:FCM1}\\
    \text{where},~{\bm i}&:=\begin{bmatrix}I_1\angle \phi_1 & I_2\angle \phi_2 &\dots &I_M\angle \phi_M\end{bmatrix}^\top, \notag\\
    {\bm v}&:=\begin{bmatrix}V_1\angle \theta_1 &V_2\angle \theta_2 &\dots &V_N\angle \theta_N\end{bmatrix}^\top, \notag\\
    v(t)&={\sum}_{h=1}^N\sqrt{2}V_h\sin\left(h\omega_0t+\theta_h\right)\,,\notag\\
    i(t)&={\sum}_{h=1}^M\sqrt{2}I_h\sin\left(h\omega_0t+\phi_h\right)\,.\notag
\end{align}
$V_h$ ($I_h$) is the RMS Fourier series coefficient, and $\theta_h$ ($\phi_h$) is the phase angle corresponding to the $h$-th harmonic components of the voltage (current) signals, while $\omega_0$ is the fundamental frequency. ${\bm i_0}\!\in\!\mathbb{C}^{M}$ is the voltage-independent component of the current harmonics, and ${\bm Y}\!\in\!\mathbb{C}^{M\times N}$ is the FCM. 

The \textit{total harmonic distortion} (THD) for current is calculated as $\sqrt{\sum_{h=2}^MI_h^2/I_1^2}$. Voltage THD is defined similarly. Moreover, the total active power (averaged over the fundamental cycle) is expressed as the sum of the active powers of the individual harmonic components as follows, \cite{grady2012understanding}:
\vspace{-1mm}
\begin{align}\label{eq:power}
    P_\text{avg}=\sum_{h=1}^{\min\left(M,N\right)}P_h,\quad P_h:=V_hI_h\cos\left(\theta_h-\phi_h\right)\,\forall h\,.
    \vspace{-1mm}
\end{align}
\textbf{FCM Estimation.}
%
As a first step to estimating FCM, we adopt the \textit{estimating signal parameters via rotational invariance techniques} \cite{roy1989esprit} method to extract the current and voltage harmonic components from their time-series waveform data. 
%
%
Given $K$ pairs of voltage and current waveform measurements, associated harmonics, $\left\lbrace\bm{i}^{(k)},\bm{v}^{(k)}\right\rbrace_{k=1}^K$, are extracted and plugged into \eqref{eq:FCM1}. After column-wise stacking those equations, we derive the following \textit{least squares} (LS) estimate of $\bm{i_0}$ and $\bm{Y}$:
\vspace{-1 mm}
\begin{equation}
    \text{(FCM Estimate)}\qquad \begin{bmatrix} \widehat{\bm i}_0 & \widehat{\bm{Y}} \end{bmatrix} = 
    \bm{I} \bm Z^+\,,
    \vspace{-0 mm}
\end{equation}
where $\bm{I} := [\bm{i}^{(1)}, \cdots, \bm{i}^{(K)}]$\,, $\bm{V} := [\bm{v}^{(1)}, \cdots, \bm{v}^{(K)}]$\,, $\bm Z := [\bm 1 \quad \bm{V}^\top]^\top$\,, and $(\cdot)^+$ denotes the pseudo-inverse of a matrix.

\textsc{Identifiability under Limited Data:} To get a meaningful estimate, the matrix $\bm{Z}$ should be well conditioned, which requires a large number of linearly independent measurements. 
In practice, though, the number of available measurements is usually limited.
Further, it is a good practice to set aside some measurements for testing, which leaves even fewer measurements for training.
This problem can be alleviated by considering only the $N'\!\leq\!N$ most significant voltage harmonics, determined by their Fourier series coefficients.
A smaller $N'$ requires fewer measurements to construct a well-conditioned matrix $\bm Z$ and, in turn, reliable estimates of FCM. 
However, this may degrade the model's capability to faithfully represent the reality.
Therefore, such a trade-off requires careful selection of $N'$ to achieve good modeling accuracy.


\section{End-use Load Modeling } \label{sec:enduse_modeling}
This section describes the chosen end-use load models. These are modeled in PSCAD tool for transient simulations.

\begin{figure}
    \centering
    \includegraphics[width=1\columnwidth]{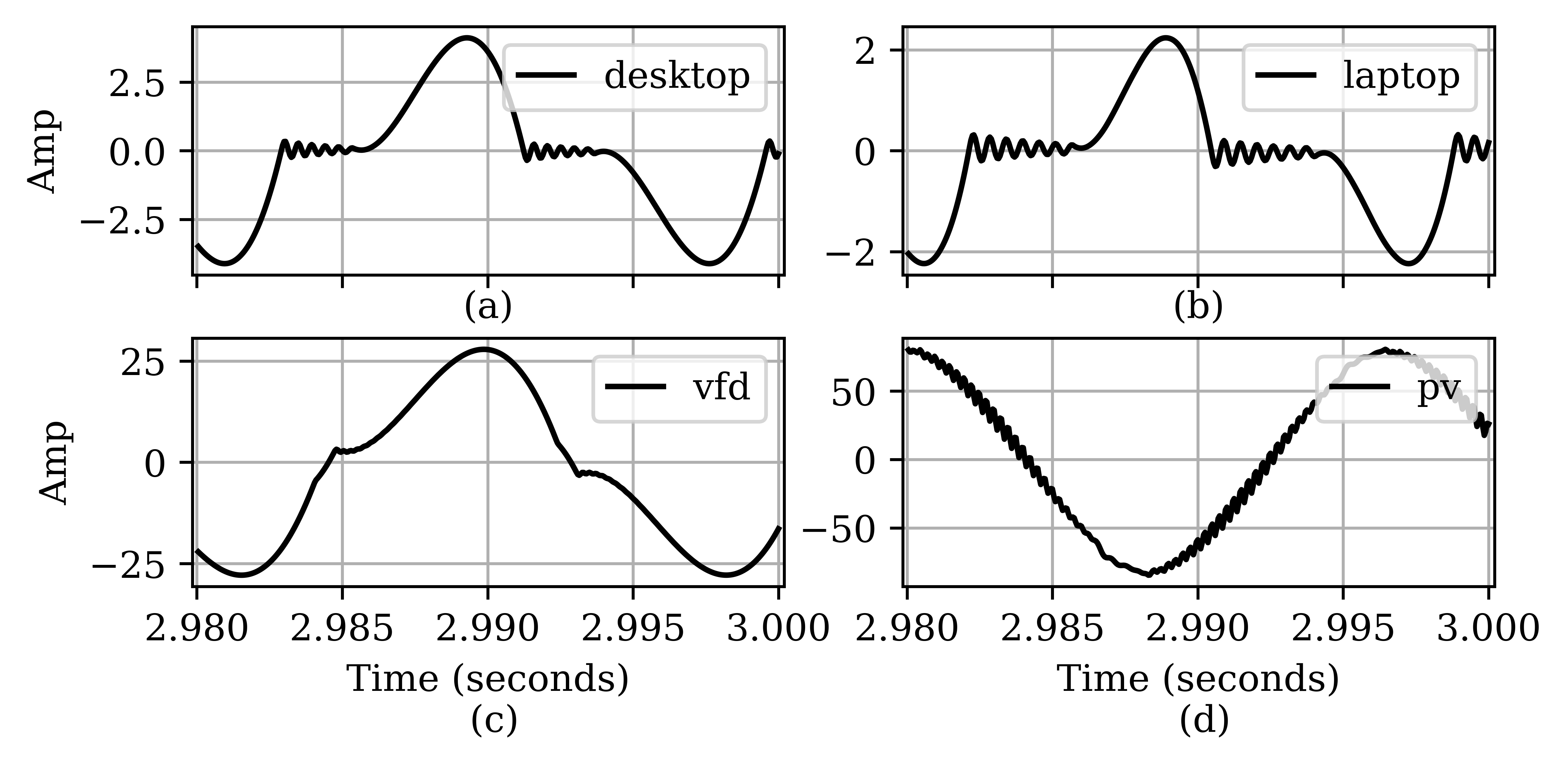}
    \vspace{-7mm}
    \caption{Input current waveform drawn by various appliances: (a) desktop, (b) laptop, (c) VFD, and (d) PV system}
    \label{fig:device_current}
    \vspace{-4mm}
\end{figure}

\textbf{Desktop Power Supply.}
A desktop computer uses a switch mode power supply (SMPS) configuration, i.e., a full-wave rectifier-based power supply followed by an appropriate DC--DC converter (e.g., buck converter), as shown in the schematic in \figurename \ref{fig:desktop_supply}. 
In this study, a simplified 200\,W desktop power supply is modeled with one buck converter added after a full-wave rectifier to get a low voltage (10\,V) supply. 
It is worth noting that the switch in the buck converter is modulated via a 6\,kHz pulse that introduces high-frequency harmonics. 
As with commercial desktops, an input filter of $(L_f, C_f, R_f)=(12 mH, 0.22 \mu F, 550 k \Omega)$ is addded to reduce the harmonics in the main supply. \figurename \ref{fig:device_current}a shows the input current waveforms of the desktop power supply with around 36\% THD.

\textbf{Laptop Charger.}
A typical laptop charger also uses an SMPS power supply that includes an input filter followed by a full-wave rectifier and a flyback DC-DC converter as shown in the schematic in \figurename \ref{fig:laptop_supply}. A 900-Hz pulse is used to drive the flyback converter switch. Input filter values used are $(L_f, C_f, R_f)=(12 mH, 0.33 \mu F, 4000 k \Omega)$. \figurename \ref{fig:device_current}b shows the steady-state input current waveform consumed by the laptop charger with around 60\% THD.

\textbf{VFD-Based Induction Motors.}
Due to the availability of only 1-phase supply in a typical residence, a 1-phase VFD is used to drive a 3-phase motor. Note that driving a 3-phase VFD from one phase will create harmonics that are speed-dependent and load-dependent. However, this study assumes that the VFD operates in a finite number of operational states, and each state (i.e., speed/frequency) represents a load profile. This is important because the developed FCM algorithm requires the dataset for a static load combination.
%
\figurename \ref{fig:vfd} shows the schematic of a 1-phase VFD-driven induction motor with a full-wave rectifier, a DC-link block, and a 3-phase DC--AC inverter model driven by a space vector modulation (SVM) control scheme \cite{bose2002}. A speed control for the VFD from \cite{bose2009} is implemented. 
%
An input filter circuit  at the grid-connected terminal with $(L_f,C_f,R_f)=(12 mH, 0.22 \mu F, 550 k \Omega)$ is chosen such that the supply current THD does not exceed 25\% as with most commercial 1-phase VFDs. 
\figurename \ref{fig:device_current}c shows the steady-state input current waveform drawn by a VFD-driven 3\,kVA induction motor.



\textbf{PV System.}
A schematic of a complete solar PV system is shown in \figurename \ref{fig:pv}. It consists of a solar panel followed by a DC--DC boost converter driven by a 3.6-kHz pulse from a \textit{maximum-power-point-tracking} (MPPT) controller. Then the boosted DC voltage is converted to AC via a single-phase DC--AC inverter. Since most commercial PV inverter systems guarantee to limit the supply current THD within 5\%, an AC filter is added with  $(L_f,C_f,R_f)=(0.9 mH, 12 \mu F, 550 k \Omega)$. \figurename \ref{fig:device_current}d shows the steady-state current waveform at the AC connection point of a 5-kW PV system with around 5\% THD. 

\section{Results and Discussion}\label{sec:results}
\begin{figure}
    \centering
    \includegraphics[width=1.0\columnwidth]{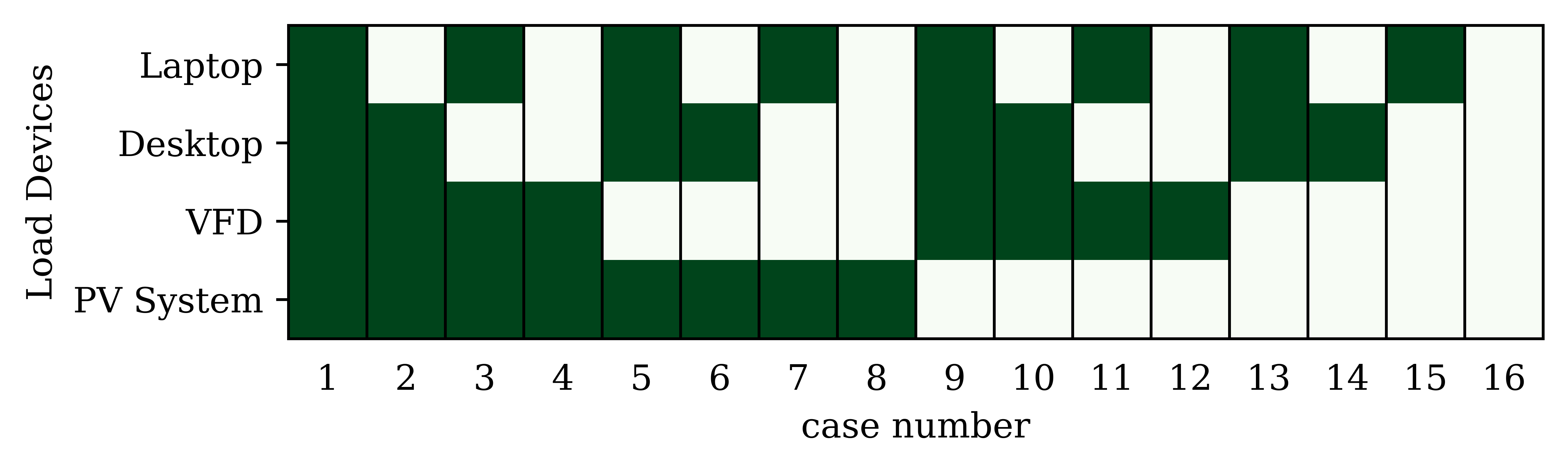}
    \vspace{-6mm}
    \caption{Map of load devices combinations to case numbers }
    \label{fig:load_map}
    \vspace{-4mm}
\end{figure}
\begin{figure}
    \vspace{-7mm}
    \centering
    \includegraphics[trim=0 0 0 3mm, clip, width=0.95\columnwidth]{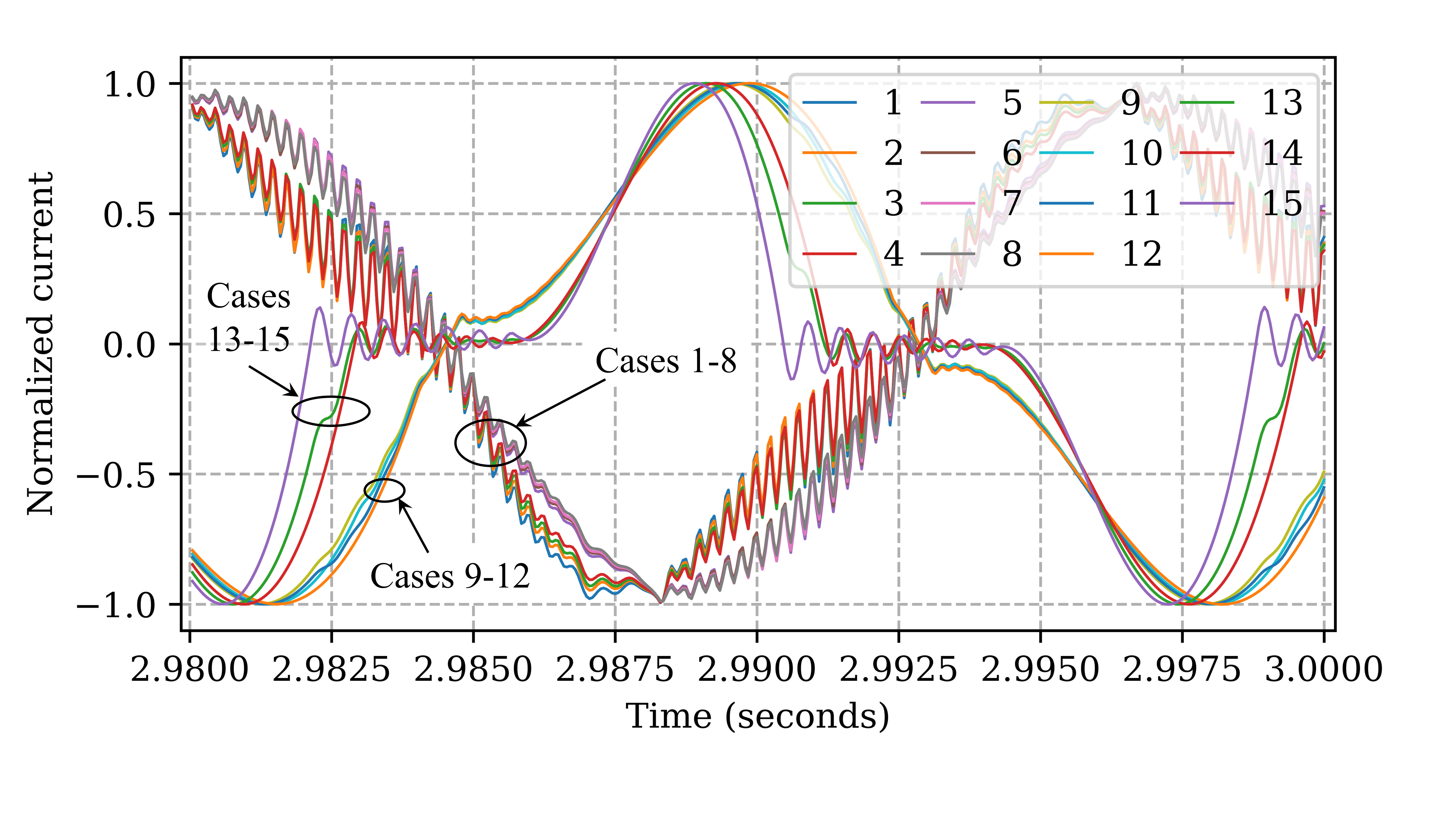}
    \vspace{-4mm}
    \caption{Normalized current ($I_s^{eq}$) waveform in all load combination cases to demonstrate the harmonic content }
    \label{fig:norm_cur_16}
\end{figure}
%
Based on the proposed process in the \figurename \ref{fig:overall_process}, 16 load combinations (cases) from the 4 load types are created as shown in \figurename \ref{fig:load_map}, and each case represents the load pattern at a particular time in a day. \figurename \ref{fig:norm_cur_16} shows the normalized supply current waveform in all 16 cases to demonstrate the harmonic content. Observing the current waveform in all cases, the following 3 groups can be made based on similar harmonic characteristics: (i) all case with PV (1--8), (ii) all cases without PV but with VFD (9--12) and, (iii) all cases without PV and VFD (13--16). For brevity, one case from each group (Cases 1, 9, 13) is picked for data-driven load modeling. 
\begin{figure}
    \centering
    \includegraphics[width=1.0\columnwidth]{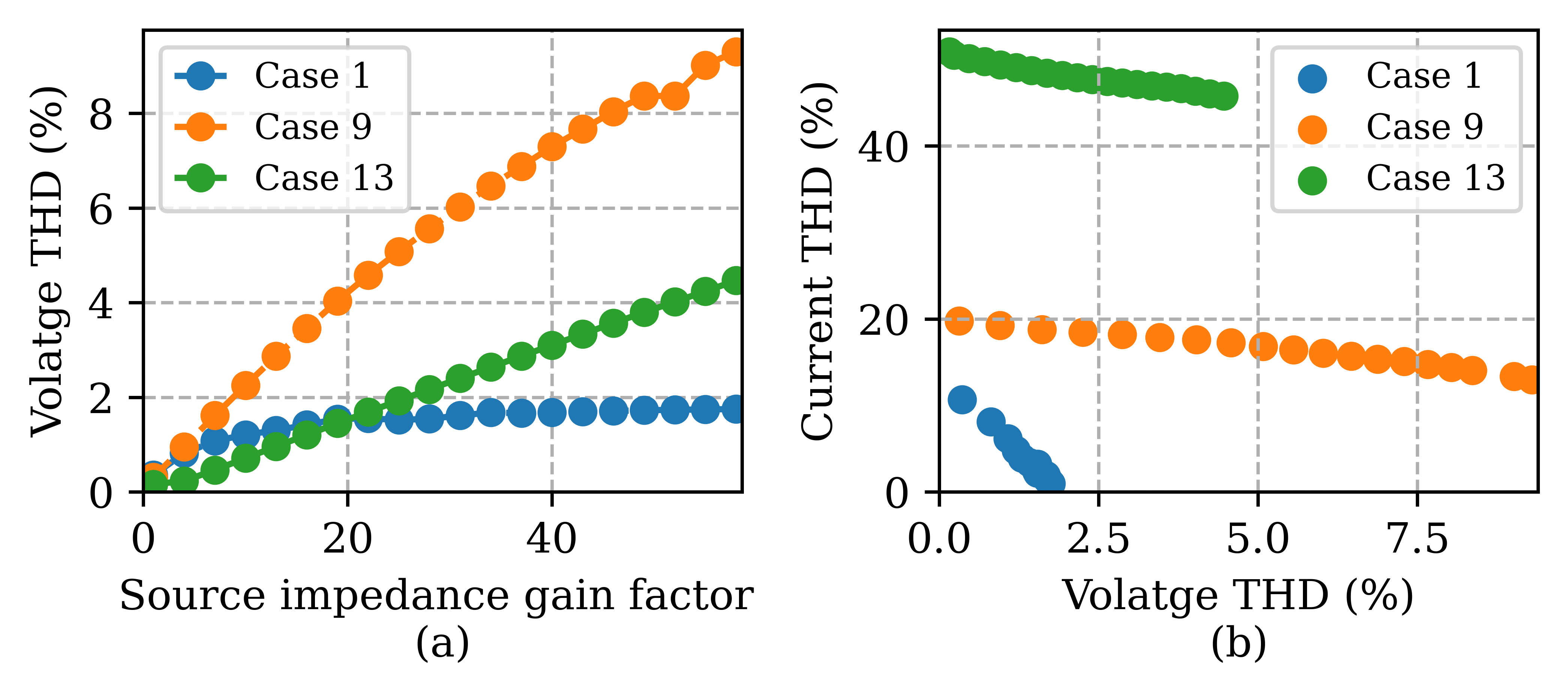}
    \vspace{-6mm}
    \caption{Impact of source impedance gain factor, $\alpha_{imp}$ on the harmonic content of supply voltage.}
    \label{fig:volt_thd}
    \vspace{-7mm}
\end{figure}
%
As discussed earlier, the impedance gain factor, $\alpha_{imp}$, is varied to model the voltage drop effect and introduce the harmonics in supply voltage, as shown in \figurename \ref{fig:volt_thd}a.
Consequently, the impact of increasing voltage THD on the current THD is shown in \figurename \ref{fig:volt_thd}b, which clearly demonstrates the \textit{`attenuation and diversity effect'} (increase in supply voltage harmonics reducing the load current harmonics), representative of real distribution networks \cite{mansoor_investigation_1995,grady_estimating_2002}. 

\begin{figure*}
    \centering
    \hspace{-0.25in}
    \begin{subfigure}[b]{0.35\linewidth}
        \centering
        \includegraphics[width=\textwidth]{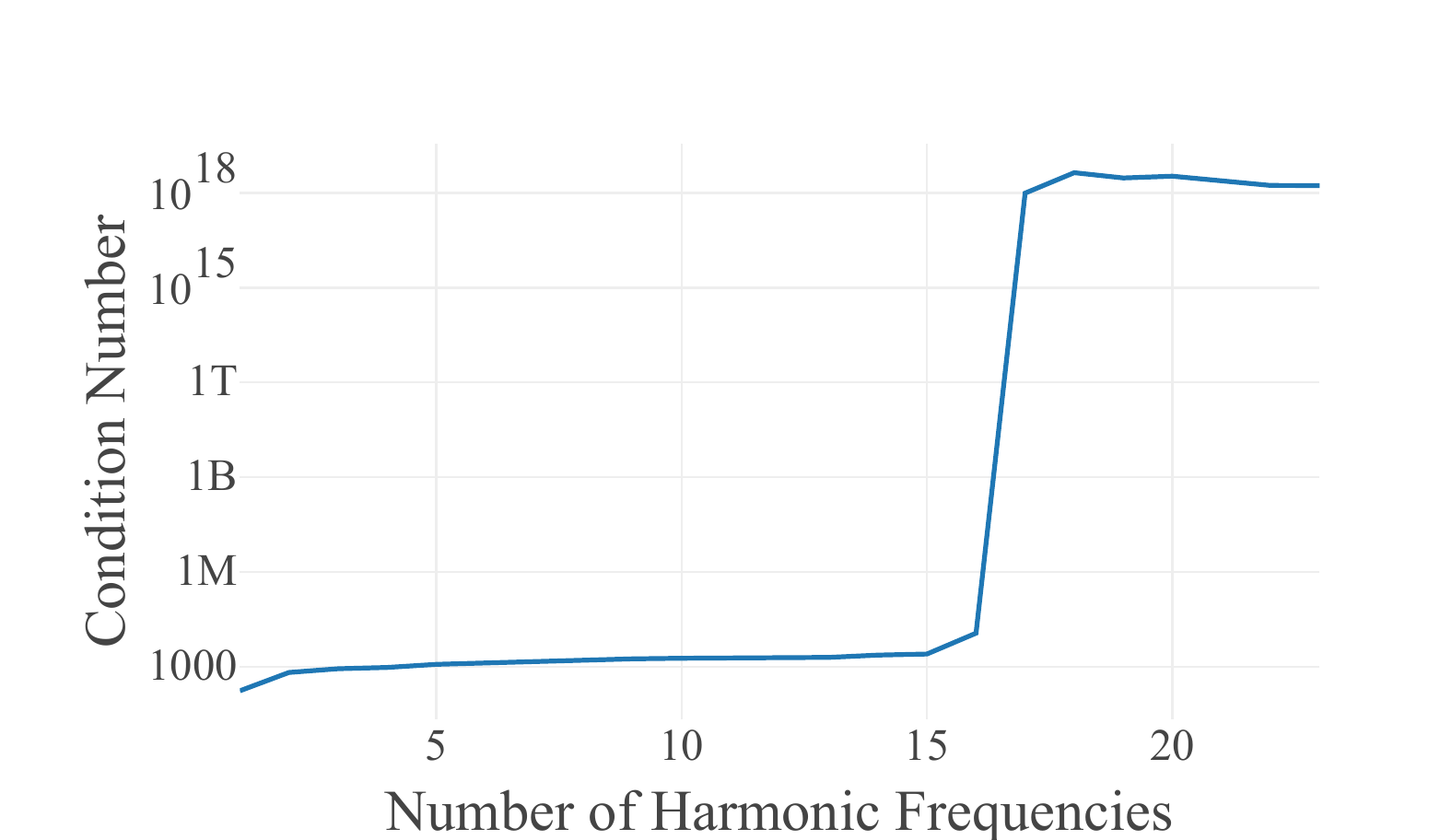}
        \caption{}
    \label{fig:case9-cond}
    \end{subfigure}
    \hspace{-0.1in}
    \begin{subfigure}[b]{0.32\linewidth}
        \centering
        \includegraphics[width=\textwidth]{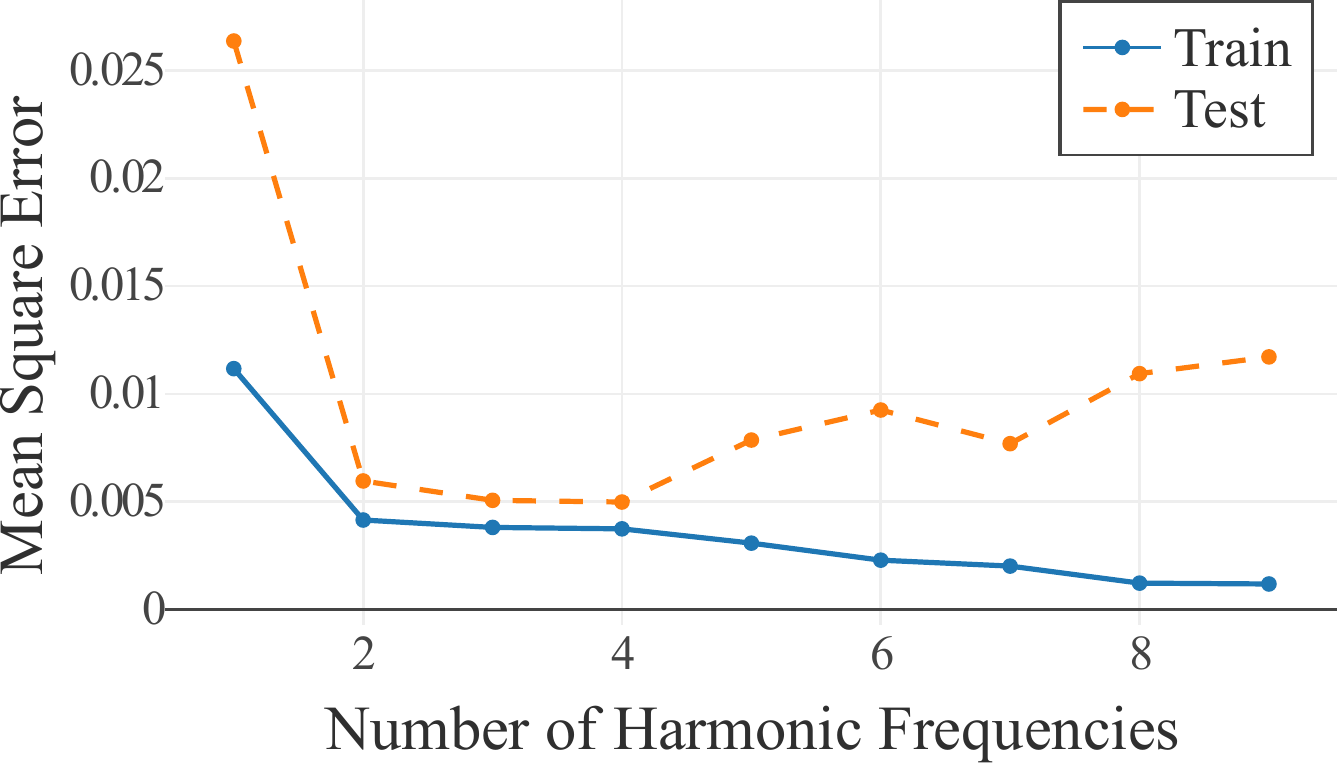}
        \caption{}
    \label{fig:case9-mse-numFreqs}
    \end{subfigure}
    \hspace{0.05in}
    \begin{subfigure}[b]{0.34\linewidth}
        \centering
        \includegraphics[width=\textwidth]{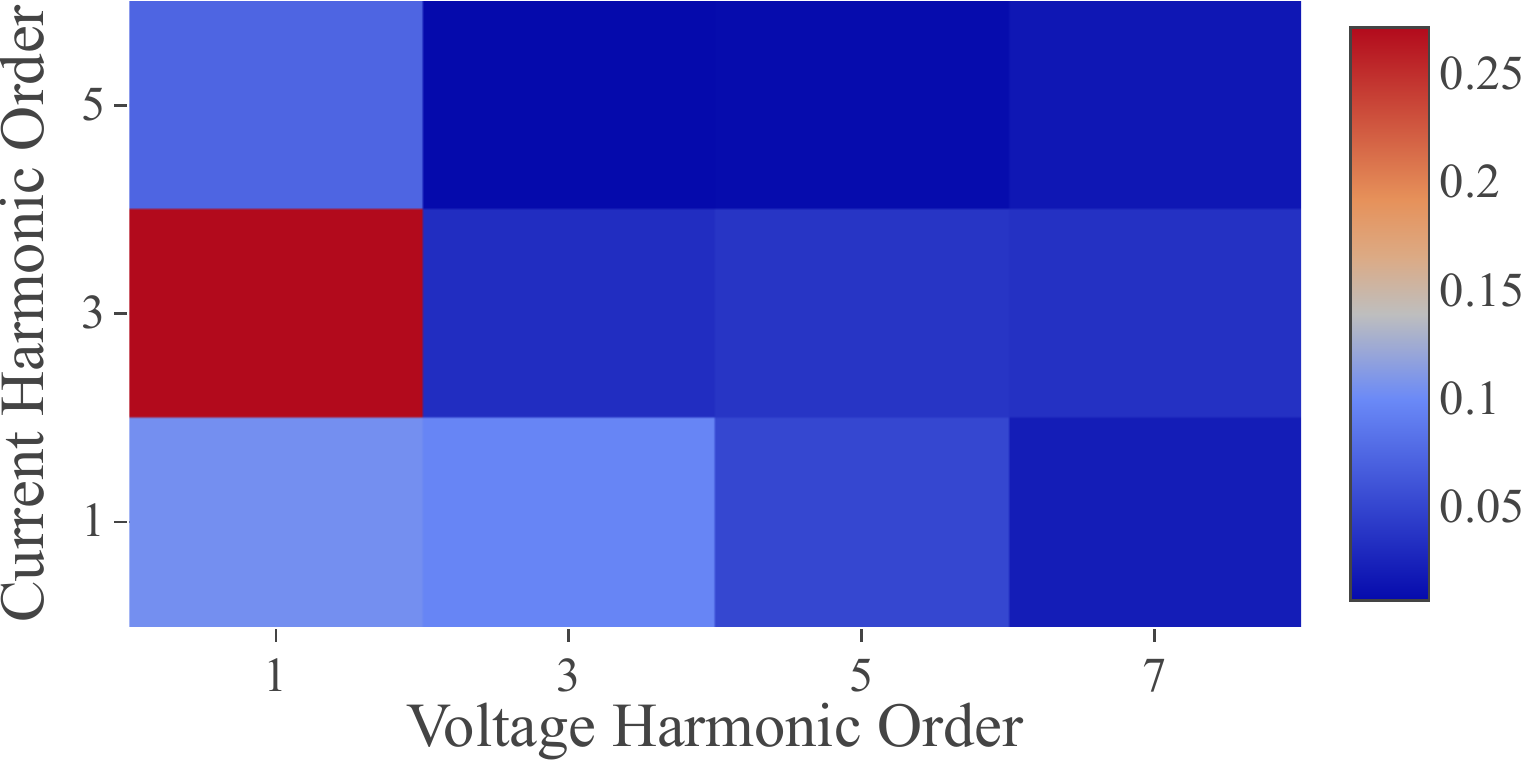}
        \caption{}
    \label{fig:case9-FCM}
    \end{subfigure}
    \hspace{-0.1in}
    \begin{subfigure}[b]{0.345\linewidth}
        \centering
        \includegraphics[width=\textwidth]{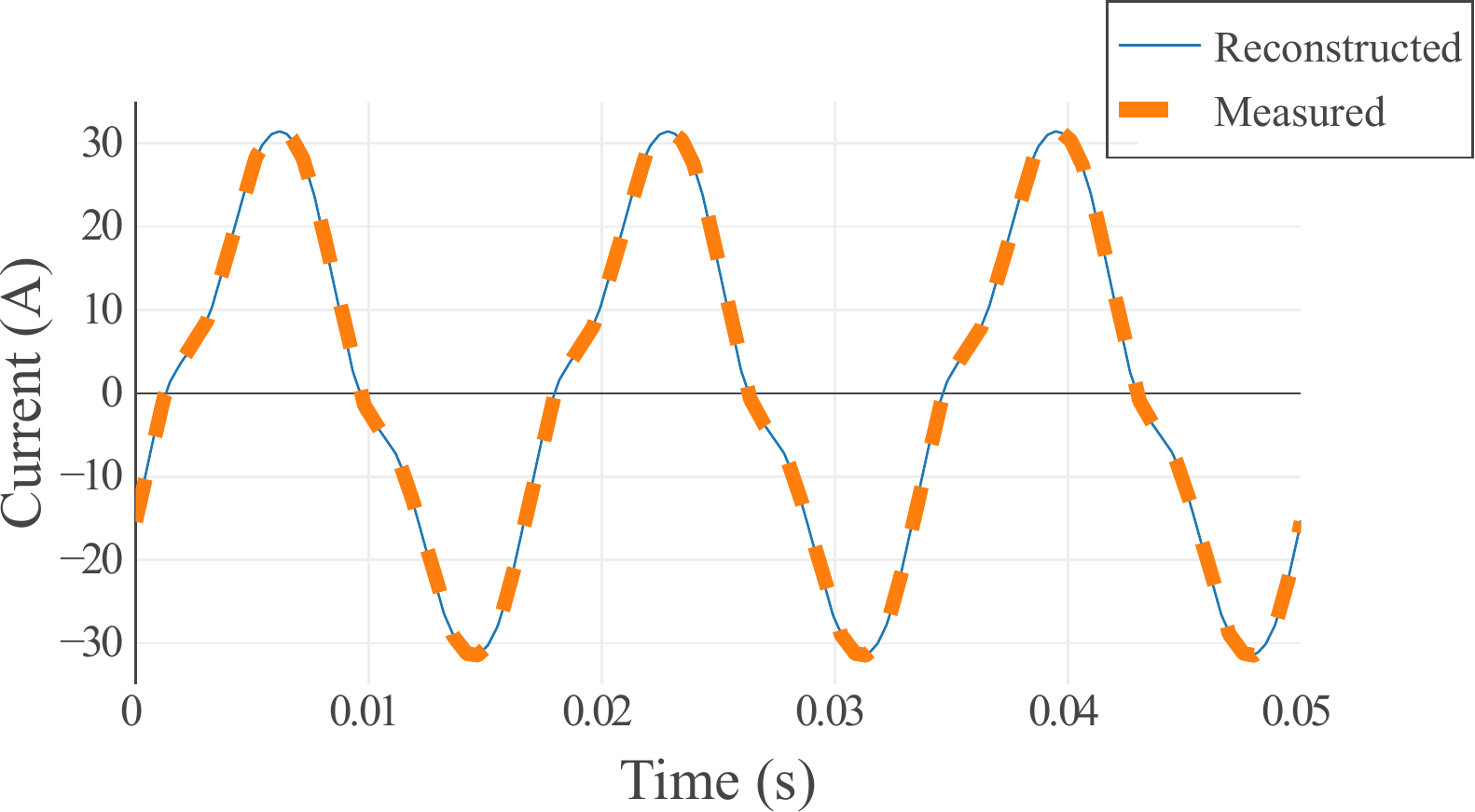}
        \caption{}
    \label{fig:case9-reconstruct}
    \end{subfigure}
    \begin{subfigure}[b]{0.305\linewidth}
        \centering
        \includegraphics[width=\textwidth]{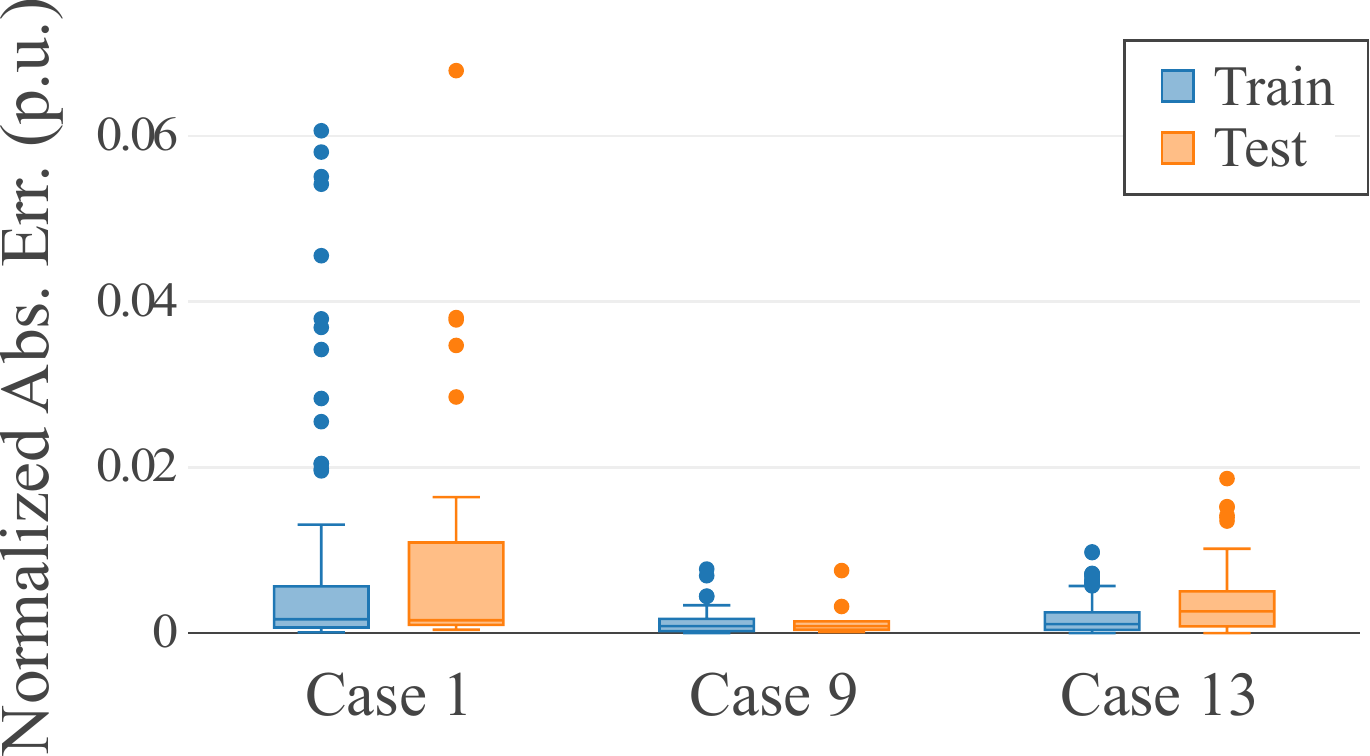}
        \caption{}
    \label{fig:errBox}
    \end{subfigure}
    \hspace{0.05in}
    \begin{subfigure}[b]{0.325\linewidth}
        \centering
        \includegraphics[width=\textwidth]{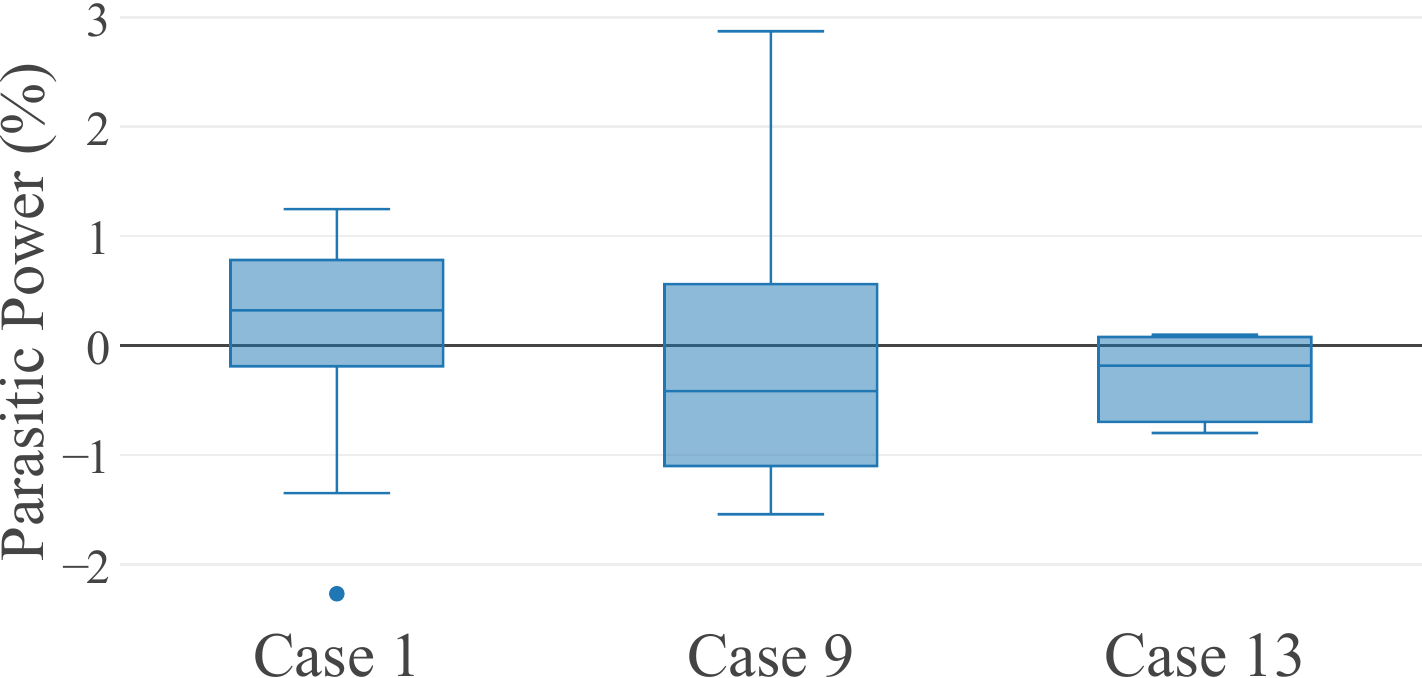}
        \caption{}
    \label{fig:thdBox}
    \end{subfigure}
    \caption{Summary results from FCM identification and performance analysis. (a) Condition number of the voltage matrix (${\bm V}$). (b) Impact of the number of selected harmonics on the estimation accuracy on training and testing data. (c) Estimated FCM (Case 9). (d) A reconstructed current signal (Case 9). (e) Reconstruction errors (all cases). (f) Parasitic power relative to fundamental.}
    \label{fig:all_FCM_figures}
    \vspace{-5mm}
\end{figure*}

In total, 19 pairs of voltage-current measurements (waveform) are generated for each load combination (Cases 1, 9, 13), and used to construct the $\bm{V}$ and $\bm{I}$ matrices. 
The variation of the condition number of the $\bm{V}$ matrix with different number of harmonics is shown in \figurename~\ref{fig:case9-cond} considering all 19 measurements.
%
The impact of the selection of $N'$, the number of considered voltage frequencies, on modeling accuracy is also explored and the results are summarized in \figurename~\ref{fig:case9-mse-numFreqs}, with 15 measurements used for training.
{As the number of frequencies increases, both training and testing errors decrease at first.
At some point (4 in this case), the testing error begins to grow while training error remains low.}
The system becomes overfit as the capacity of the system continue to increase as a result of larger number of harmonics.
This yields poor generalization performance, emphasizing the need to chose the appropriate $N'$ for accurate FCM estimation.
%
%
%
%
The FCM estimate for Case 9 is shown in \figurename~\ref{fig:case9-FCM}. {It can be seen, in Case 9, the 1st (fundamental) and 3rd harmonics in voltage contributed significantly to the fundamental frequency in current. The 3rd and 5th harmonics in current are mainly the artifacts of the fundamental frequency in voltage.}
%
%
%
\figurename~\ref{fig:case9-reconstruct} illustrates how the reconstructed current signal (using the estimated FCM) closely matches the measured current waveform. The overall performance of the FCM in reconstructing current waveforms across the three cases is presented in \figurename~\ref{fig:errBox}, in the form of a box plot of the \textit{normalized absolute errors} calculated by $\left|\,(\widehat{\bm{i}}_0 + \widehat{\bm{Y}} \bm{v}^{(k)} - \bm{i}^{(k)})/I_1^{(k)}\,\right|\,\forall k\!=\!1,2,\dots,K$\,.
%
%
This demonstrates the robust accuracy of the FCM estimates, with a vast majority of errors across all cases lying below 1\%. Finally, the expression in \eqref{eq:power} is used to compute the aggregated harmonic (\textit{parasitic}) power  $P_2+\dots+P_{\min(M,N)}$, and plot its value relative to the fundamental component ($P_1$) in Fig.\,\ref{fig:thdBox}. Note that the fundamental component also denotes the power estimated by the conventional ZIP modeling. The harmonic (parasitic) power is seen to be up to 2-3\% of the fundamental component, a significant contributor to the total system losses \cite{grady_estimating_2002}, that could not be captured by the ZIP modeling method.


\vspace{-0.5mm}
\section{Conclusion}\label{sec:conclusion}
\vspace{-0.5mm}
This work presents a systematic procedure to generate realistic data sets for harmonic load modeling of an electronics-dominated residence via EMTP simulations. The proposed procedure is built on the detailed modeling of various individual power electronics appliances. Further, an impedance gain technique is proposed to model the interaction of these loads, i.e., the propagation of load current harmonics into the supply voltage and consequently, the impact of distorted voltage supply on the load current harmonics. Finally, an improved data-driven FCM methodology for harmonic-enhanced load modeling is described, and its accuracy is demonstrated on data generated using the proposed procedure. 
Future work will focus on development of online identification methods that continually adapt to varying load compositions, validation using real datasets, and system-level impact analysis of harmonics.
\vspace{-1mm}
\section*{Acknowledgment}
\vspace{-1mm}
This work was supported by the Sensors and Data Analytics Program of the U.S. Department of Energy Office of Electricity, under Contract No. DE-AC05-76RL01830.
\ifCLASSOPTIONcaptionsoff
  \newpage
\fi
\bibliographystyle{IEEEtran}
\vspace{-1mm}
\bibliography{./bibtex/references_HELM}

\begin{thebibliography}{10}
\providecommand{\url}[1]{#1}
\csname url@samestyle\endcsname
\providecommand{\newblock}{\relax}
\providecommand{\bibinfo}[2]{#2}
\providecommand{\BIBentrySTDinterwordspacing}{\spaceskip=0pt\relax}
\providecommand{\BIBentryALTinterwordstretchfactor}{4}
\providecommand{\BIBentryALTinterwordspacing}{\spaceskip=\fontdimen2\font plus
\BIBentryALTinterwordstretchfactor\fontdimen3\font minus
  \fontdimen4\font\relax}
\providecommand{\BIBforeignlanguage}[2]{{%
\expandafter\ifx\csname l@#1\endcsname\relax
\typeout{** WARNING: IEEEtran.bst: No hyphenation pattern has been}%
\typeout{** loaded for the language `#1'. Using the pattern for}%
\typeout{** the default language instead.}%
\else
\language=\csname l@#1\endcsname
\fi
#2}}
\providecommand{\BIBdecl}{\relax}
\BIBdecl

\bibitem{mclorn2017enhanced}
G.~McLorn, J.~Morrow, D.~Laverty, R.~Best, X.~Liu, and S.~McLoone, ``{Enhanced
  ZIP load modelling for the analysis of harmonic distortion under Conservation
  Voltage Reduction},'' \emph{CIRED-Open Access Proceedings Journal}, vol.
  2017, no.~1, pp. 1094--1097, 2017.

\bibitem{grady2012understanding}
M.~Grady, ``Understanding power system harmonics,'' \emph{Austin, TX:
  University of Texas}, 2012.

\bibitem{Brunoro2017}
M.~Brunoro, L.~F. Encarna{\c{c}}{\~a}o, and J.~F. Fardin, ``Modeling of loads
  dependent on harmonic voltages,'' \emph{Electric power systems research},
  vol. 152, pp. 367--376, 2017.

\bibitem{collin_component-based_2010}
A.~J. Collin, J.~L. Acosta, B.~P. Hayes, and S.~Z. Djokic, ``Component-based
  aggregate load models for combined power flow and harmonic analysis,'' in
  \emph{7th {Med-Power}}, Nov. 2010, pp. 1--10.

\bibitem{Salles_2012}
D.~Salles, C.~Jiang, W.~Xu, W.~Freitas, and H.~E. Mazin, ``{Assessing the
  Collective Harmonic Impact of Modern Residential Loads—Part I:
  Methodology},'' \emph{IEEE Trans Pow Del}, vol.~27, no.~4, pp. 1937--1946,
  2012.

\bibitem{patidar_harmonics_2009}
R.~D. Patidar and S.~P. Singh, ``Harmonics estimation and modeling of
  residential and commercial loads,'' in \emph{2009 {International}
  {Conference} on {Power} {Systems}}, Dec. 2009, pp. 1--6.

\bibitem{mansoor_investigation_1995}
A.~Mansoor, W.~Grady, A.~Chowdhury, and M.~Samotyi, ``An investigation of
  harmonics attenuation and diversity among distributed single-phase power
  electronic loads,'' \emph{IEEE Trans Power Del}, vol.~10, no.~1, pp.
  467--473, Jan. 1995.

\bibitem{grady_estimating_2002}
W.~Grady, A.~Mansoor, E.~Fuchs, P.~Verde, and M.~Doyle, ``Estimating the net
  harmonic currents produced by selected distributed single-phase loads:
  computers, televisions, and incandescent light dimmers,'' in \emph{{IEEE PES
  Winter Meeting}}, vol.~2, Jan. 2002, pp. 1090--1094 vol.2.

\bibitem{zhao2004harmonic}
Y.~Zhao, J.~Li, and D.~Xia, ``Harmonic source identification and current
  separation in distribution systems,'' \emph{International journal of
  electrical power \& energy systems}, vol.~26, no.~1, pp. 1--7, 2004.

\bibitem{Fauri_1997}
M.~Fauri, ``Harmonic modelling of non-linear load by means of crossed frequency
  admittance matrix,'' \emph{IEEE Transactions on Power Systems}, vol.~12,
  no.~4, pp. 1632--1638, 1997.

\bibitem{lehn2007frequency}
P.~Lehn and K.~Lian, ``Frequency coupling matrix of a voltage-source converter
  derived from piecewise linear differential equations,'' \emph{IEEE Trans
  Power Del}, vol.~22, no.~3, pp. 1603--1612, 2007.

\bibitem{malagon-carvajal_harmonic_2020}
G.~Malagon-Carvajal, C.~Duarte, G.~Ordoñez-Plata, C.~F.~M. Almeida, and
  N.~Kagan, ``A {Harmonic} {Frequency} {Domain} {Modeling} {Method} for
  {Single}-{Phase} {Full} {Bridge} {Rectifiers},'' \emph{Journal of Control,
  Automation and Electrical Systems}, vol.~31, no.~5, pp. 1322--1333, Oct.
  2020.

\bibitem{lennerhag_stochastic_2020}
O.~Lennerhag and M.~H.~J. Bollen, ``A {Stochastic} {Aggregate} {Harmonic}
  {Load} {Model},'' \emph{IEEE Trans on Power Delivery}, vol.~35, no.~5, pp.
  2127--2135, Oct. 2020.

\bibitem{senra2017assessment}
R.~Senra, W.~C. Boaventura, and E.~M. Mendes, ``Assessment of the harmonic
  currents generated by single-phase nonlinear loads,'' \emph{Electric power
  systems research}, vol. 147, pp. 272--279, 2017.

\bibitem{roy1989esprit}
R.~Roy and T.~Kailath, ``{ESPRIT-estimation of signal parameters via rotational
  invariance techniques},'' \emph{IEEE Transactions on acoustics, speech, and
  signal processing}, vol.~37, no.~7, pp. 984--995, 1989.

\bibitem{bose2002}
B.~Kumar, \emph{Modern Power Electronics \& AC Drives}.\hskip 1em plus 0.5em
  minus 0.4em\relax Prentice Hall, 2002.

\bibitem{bose2009}
B.~K. Bose, ``Power electronics and motor drives recent progress and
  perspective,'' \emph{IEEE Trans Ind Electron}, vol.~56, no.~2, pp. 581--588,
  2009.

\end{thebibliography}
\end{document}